\documentclass[a4paper,11pt]{article}
\usepackage{amsthm}
\usepackage{amsmath}
\usepackage{amssymb}
\usepackage{color}
\usepackage{cite}
\usepackage{graphicx}
\usepackage{xcolor}
\usepackage{tikz}
\usetikzlibrary{decorations.pathmorphing}
\usetikzlibrary{decorations.pathreplacing}
\usepackage{caption}
\usepackage[linktocpage]{hyperref}
\usepackage{setspace}
\usepackage{titletoc}
\usepackage{soul}
\newcommand{\mrd}{\mathrm{d}}

\definecolor{darkred}{rgb}{0.9,0.05,0.05}
\definecolor{darkblue}{rgb}{0.05,0.05,0.6}
\definecolor{darkgreen}{rgb}{0.05,0.6,0.05}
\definecolor{brightgreen}{rgb}{0.1,0.9,0.1}

\renewcommand*{\eqref}[1]{%
  \begingroup
    \hypersetup{
      linkcolor=darkblue,
      linkbordercolor=darkblue,
    }%
    \textcolor{darkblue}{(\ref{#1})}%
  \endgroup 
}

\hypersetup{linkcolor=red,citecolor=brightgreen,urlcolor=black,colorlinks=true}

\makeatletter
\newcommand\footnoteref[1]{\protected@xdef\@thefnmark{\ref{#1}}\@footnotemark}
\makeatother

\begin{document}
\let\endtitlepage\relax

\begin{titlepage}
\begin{center}

\newcommand\blfootnote[1]{%
  \begingroup
  \renewcommand\thefootnote{}\footnote{#1}%
  \addtocounter{footnote}{-1}%
  \endgroup
}

\vspace*{0.0cm}

{\fontsize{19pt}{22pt}\bf{On the Origin of Black Hole Paradoxes}}
 
\vspace{8mm}
\centerline{\large{Kamal Hajian$^{\ast\dagger}$\blfootnote{$^\ast$e-mail: kamal.hajian@uni-oldenburg.de}\blfootnote{$^\dagger$e-mail: khajian@metu.edu.tr} }}
\vspace{-1mm}
\normalsize
\textit{Institut für Physik, Universität Oldenburg, Postfach 2503, D-26111
Oldenburg, Germany}\\
 \textit{Department of Physics, Middle East Technical University, 06800, Ankara, Turkey}

%--------------------------------------------------------------------------------------------------------------------------------------
\begin{abstract}
Black hole firewall paradox is an inconsistency between four postulates in black hole physics: (1) the unitary evolution in quantum systems, (2)  application of the semi-classical field theory in low curvature backgrounds, (3) statistical mechanical origin of the black hole entropy, and (4) the equivalence principle in the version of no drama for free-falling observers in the vicinity of the horizon. Based on the existence of the Hawking radiation for the static observers standing outside a Schwarzschild black hole, we show a direct contradiction between the postulates (2) and (4). If there is not a way out of this new problem, it implies the necessity of relaxing one of these two assumptions for resolving the black hole firewall paradox.

\vspace{2mm}
\end{abstract}
\end{center}
\vspace*{0cm}
\end{titlepage}
\vspace*{-4mm}

\renewcommand{\baselinestretch}{1.0}  %Line spacing
\setstretch{1}

\section{Introduction}
The semi-classical regime is studying quantum fields on a classical background gravity. Although a black hole (BH) is classically the most perfect black object, at the semi-classical regime it emits thermal radiation  which is called Hawking radiation \cite{Hawking:1974rv,Hawking:1976rt}.  In the early 70s, theoretical studies of BHs revealed that they behave similar to a thermodynamic system. Specifically, they have entropy \cite{Bekenstein:1973ft}, temperature \cite{Hawking:1976rt}, and satisfy the four laws of thermodynamics \cite{Bardeen:1973gd}.  Accordingly, the Hawking radiation is analogous to the thermal radiation which is emitted from a thermodynamic system. However, in 1976 Stephen Hawking realized that as far his calculation shows, there is an unexpected difference between these radiations: the process of emission of the Hawking radiation breaks the unitarity of the evolution at the semi-classical regime \cite{Hawking:1976ra} if the evaporation of the BH is taken into account. This is in contrast with the thermal radiation from other objects which respects unitarity in the context of the standard quantum field theory. This persisting inconsistency is called BH information paradox.   

Since then, there have been many proposals to resolve the information paradox  (see reviews in \cite{Mathur:2009hf,Bultinck:2013yga,Carlip:2014pma,Lust:2018cvp}). The majority of the resolutions are studied on a Schwarzschild BH (either an eternal or a collapsing one) for simplicity, and we follow this simplification in this paper. One of the suggested resolutions is called BH complementarity which was proposed in 1993 \cite{Susskind:1993if}.  The idea of complementarity is based on the membrane paradigm for modeling the BH horizon. In this model,  the horizon is described as a hot membrane (also called stretched horizon) for observers which are fixed outside the BH and in the vicinity of the horizon. This is in agreement with the blue-shift of the Hawking radiation as seen by the fiducial observers (Fido \cite{note1}), i.e. static observers which are fixed outside of the horizon at a constant radius $r_0$.  As the Hawking radiation is followed back towards its origin around the BH horizon, it is perceived locally by the Fidos  to be  hotter and hotter until one reaches approximately a Planck distance $\ell_p$ outside the horizon. From the point of view of the Fidos, here there is a stretched horizon  which is considered to be the source of emission of the Hawking radiation. In the semi-classical regime, replacement of the BH interior by such a hot membrane could basically remedy the unitarity problem.  

However, in the BH complementarity a question arises: what could a free-falling observer (Frefo \cite{note1}) which dives into the BH perceive from the stretched horizon?  In the BH complementarity, the answer is: nothing dramatic. In other words, a Frefo passes through the horizon without hitting a hot surface. This picture is supported by the equivalence principle, which states that in the low-curvature regions (such as around the horizon of a large enough BH) the physics of a Frefo is described locally as an inertial observer in the flat space-time. In this regard, in the BH complementarity the Fidos have the stretched horizon and a unitary evolution, while Frefos diving into the horizon do not hit such a membrane. Their physics  could be complementary to each other: as seen by a Fido, {a particle which is sent towards a black hole is absorbed in the stretched horizon, thermalized, and returns in Hawking like radiation,} while as seen by a Frefo the particle passes through the horizon and enters the black hole.    

BH complementarity was considered as one of the successful resolutions for the information paradox. However, in a seminal paper \cite{Almheiri:2012rt} in 2012 it was shown that there is an inconsistency between the assumptions of the complementarity. This inconsistency is known as the ``firewall paradox." In order to have a self contained paper, we review the paradox briefly here (more details can be found in the original paper \cite{Almheiri:2012rt} or reviews e.g. in \cite{Tycho Sikkenk,Dundar:2014gma,Mann:2015luq,Polchinski:2016hrw}). The highlighted assumptions in the firewall paradox are as follows.

\begin{itemize}
\item[] \textit{Postulate 1:} 
 The process of formation and evaporation of a black hole, as viewed by a distant observer, can be described entirely within the context of standard quantum theory. In particular there exists a unitary S-matrix which describes the evolution from infalling matter to outgoing Hawking-like radiation.

\item[] \textit{Postulate 2:} Outside the stretched horizon of a massive black hole, physics can be described to good approximation by a set of semi-classical field equations.

\item[] \textit{Postulate 3:} To a distant observer, a black hole appears to be a quantum system with discrete energy levels. The dimension of the subspace of states describing a black hole of mass M is the exponential of the Bekenstein entropy $S(M)$.

\item[] \textit{Postulate 4:} A freely falling observer experiences nothing out of the ordinary when crossing the horizon.
\end{itemize}
In the literature, these four assumptions might be referred to as the Unitarity, Semi-classical Approximation, Central Dogma, and No Drama respectively, and they are the basic assumptions in the BH complementarity. The authors of the firewall paper provided a setup to show an inconsistency between these assumptions. They focused on a black hole after its Page time and called it an old black hole. The Page time is the time when the black hole loses  approximately half of its entropy via the Hawking radiation. If the BH radiation process is unitary, then after the Page time the entanglement entropy of the radiated particles with the BH decreases by time \cite{Page:1993df,Page:1993wv}. In other words, after the Page time the earlier radiated system of particles is purified by the later radiated particles. 

In the firewall paradox setup, one focuses on a typical moment of the old ages of a BH. The radiation that this old BH has already emitted is denoted by $R$.  The Hawking particle which is emitted at the chosen time is named as the particle $B$. Because the Hawking particles are created in pairs, the particle $B$ has an infalling partner towards the inside of the BH, which is called the particle $A$. The particle $A$ and particle $B$ must be maximally entangled, because the state that they create together should be the vacuum with respect to the infalling observers. This is required by the equivalence principle: Postulate 4. On the other hand, the particle  $B$ is maximally entangled with a subsystem of the early radiation $R$. This is required by Page argument, which is based on the unitarity: Postulate 1. However,  the mutual maximal entanglements between $A$--$B$ and $B$--$R$ are not allowed (this is called monogamy) because of the strong sub-additivity of entanglement entropies in quantum systems. Hence one obtains a paradox. We notice that the existence of the Hawking radiation is based on the semi-classical approximation: Postulate 2. Besides, the Postulate 3 is implicitly used in the Page argument in order to prevent extra degrees of freedom for an old BH. We recall that the Page time is defined in terms of the degrees of freedom of a BH (via its entropy), and this postulate does not allow the Page time to be postponed arbitrarily by introduction of hypothetical extra degrees of freedom. Therefore, the firewall paradox setup uses all of the 4 postulates to show an inconsistency between them. 

Which one of the four postulates is the problematic one? This a question which arises, and can be considered as a continuation of the information loss puzzle from 1976. In this regard, the history of proposing resolutions to this question is as old as the information paradox itself. There have been many different attempts to find the problem. As an example,  the authors in the firewall paper \cite{Almheiri:2012rt}  suggest the Postulate 4 as the problematic one, and relax it by introducing a wall of fire at the horizon which prevents a Frefo to fall into the BH. The terminology of the ``firewall" comes from this suggested resolution. For a non-exclusive reference of  other recent approaches and arguments to find the origin of the problem see the papers \cite{Susskind:2012rm,Almheiri:2013hfa,Avery:2012tf,Verlinde:2013uja,Verlinde:2013vja,Giddings:2012bm,Giddings:2012gc,Giddings:2013kcj,Smerlak:2013sga,Papadodimas:2012aq,Papadodimas:2013jku,Papadodimas:2013wnh,Harlow:2014yoa,Bousso:2012as,Page:2013mqa,Harlow:2013tf,Hutchinson:2013kka,Maldacena:2013xja,Ilgin:2013iba,Mathur:2013gua,Brustein:2013qma,Brustein:2013uoa,Susskind:2014yaa,Rovelli:2014cta,Chen:2014jwq,Nomura:2014voa,Braunstein:2014nwa,Perez:2014xca,Modak:2014vya,Nikolic:2014mga,zumFelde:2014ahc,Gim:2014swa,  Hawking:2015qqa,Dundar:2015axd,tHooft:2015pce,Mersini-Houghton:2015yxx,Modak:2016uwr,Barbado:2016nfy,Unruh:2017uaw,Rovelli:2019tbl,Almheiri:2020cfm,Gan:2020nff,Perry:2021udd,Buoninfante:2021ijy}. With a quick search through the literature, one can be convinced that the majority of BH physicists do not agree on the origin of the information and firewall paradoxes. 

In this regard, it can be helpful to make the set of contradicting assumptions as small as possible. Such a contraction of the set of inconsistent postulates can provide a criterion to discard some of the proposals which take the inconsistent assumptions as granted.  In this paper, we try to suggest such an analysis to make the set of contradicting assumptions smaller. The final result of our analysis shows that the Postulate 2 (i.e. the semi-classical approximation) and the Postulate 4 (i.e. the No Drama) are in contradiction with each other, and one of them should be relaxed. 

To give an overview to the reader, we clarify the logic behind the analysis in this paper.
\begin{enumerate}
\item Fidos observe hot thermal radiation emanating from the vicinity of the horizon and propagating outward.
\item Due to standard thermodynamics, thermal radiation exerts pressure on the objects; the hotter the radiation, the higher the pressure.
\item From the point of view of Fidos and due to the standard dynamics of moving objects, the Hawking radiation pressure yields a deviation of motion for free-falling objects.  
\item However, the deviation from the geodesic motion is observer-independent and can be detected by the free-falling observer herself or himself.
\end{enumerate}
{In this regard, the paradox can be summarized as: in the semi-classical regime, static observers which reside outside a black hole observe that the interaction (via the pressure) of the Hawking radiation with a free-falling object yields deviations in its geodesic motion, while this effect becomes more significant as the object gets closer to the black hole horizon where the Hawking radiation is hotter. But, the deviation from geodesic motion is observer independent, and the free-falling observers admit the deviation too. So, if it is considered from the point of view of the free-falling object itself, there is a deviation from the geodesic motion, and there is a tension with the equivalence principle and expectation of no radiation seen by the falling observers.}

\section{Inconsistency of the Postulates 2 and 4}
Our aim in this section is to show that, without assuming the Postulates 1 and 3 (i.e. the Unitarity and Central Dogma), the Postulates 2 and 4 (i.e. the Semi-classical Approximation and No Drama) yield an inconsistency.  An overview of what we are going to do is as follows: providing carefully some classical and semi-classical arguments,  we show that if the Hawking radiation  (partially or completely) exists for the Fidos outside a BH (which it is guaranteed by the Postulate 2) then there exists a firewall for Frefos at the vicinity of the horizon, and this is in contradiction with the No Drama postulate.  

In the rest of the paper, we will go step by step to clarify and make accurate the inconsistency which was briefly advertised. In what follows, we will try to make the main statements and assumptions of the argument as bold as possible.  For convenience and simplicity, we put $c=\hbar=G=k_B=1$ and focus on the Schwarzschild BH which is spherically symmetric.  The analysis in this paper does not depend on whether the Schwarzschild BH is eternal or formed by a collapse of matter. From the point of view of a distant observer, the outside region of a Schwarzschild BH of the mass $M$ is described in the standard spherical coordinates $(t,r,\theta,\varphi)$ by the metric 
\begin{equation}\label{Schwarzschild}
\mrd s^2=-f \mrd t^2+\frac{\mrd r^2}{f}+r^2(\mrd \theta^2+\sin^2\theta \mrd \varphi^2), \qquad f(r)=1-\frac{2M}{r}.
\end{equation}
The horizon radius $ r_{_\mathrm{H}}$ is equal to $2M$. We call an observer which is fixed at a constant radius $r_0\geq  r_{_\mathrm{H}}+\ell_p$  as ``$r_0$-Fido." The world-line of an $r_0$-Fido is parametrized by $x^\mu=(t,r_0,\theta_0,\varphi_0)$ for some unimportant $(\theta_0,\varphi_0)$ and arbitrary time $t$. However, it is implicitly assumed that the time $t$ is always large enough such that the $r_0$-Fidos can receive radiations from the vicinity of the BH. In this language, the asymptotic observer is an $r_0$-Fido with $r_0\gg r_{_\mathrm{H}}$. 
\begin{itemize}
\item [] {\textbf{Statement 1}:} At the semi-classical regime and as seen by the asymptotic observers, there exists a radially outgoing thermal radiation with the temperature $T=\frac{1}{8\pi M}$. This statement does not rely on the Postulate 1 or the Postulate 3.
\end{itemize}
The first part of the statement is the celebrated Hawking radiation, and we have taken it as granted and have highlighted it as an input of our analysis. The second part emphasizes that it is not necessary to assume the Unitarity or Central Dogma in order to derive the Hawking radiation for an asymptotic observer. There are different methods to derive the Hawking radiation, e.g.  Bogoliubov coefficients method \cite{Bogoliubov} which the original Hawking calculation is based on it \cite{Hawking:1976rt,Hawking:1976ra}, the  Parikh-Wilczek tunneling model \cite{Parikh:1999mf}, Hawking radiation from the trace anomaly of the stress-energy tensor near the horizon \cite{Davies:1976ei,Christensen:1977jc,Robinson:2005pd,Iso:2006ut,Iso:2007nf}, periodic Green's functions method \cite{Gibbons:1976es,Gibbons:1976pt}, and gravitational partition function for the Euclidean BH \cite{Gibbons:1976ue}. Investigation of these methods supports Statement 1. For a nice review of these  (and other) methods see Carlip's review paper \cite{Carlip:2014pma}, specifically its section 5.1 where the necessary assumptions of the derivation of the Hawking radiation is discussed. However, in order to have a self contained paper, the Bogoliubov method is reviewed in the Appendix to show the independence of the existence of Hawking radiation from the Postulates 1 and 3 explicitly.
\begin{itemize}
\item [] {\textbf{Statement 2}:} An observer who is suspended just above the stretched horizon, he or she will observe an intense flux of
energetic radiation apparently emanating from the membrane.
\end{itemize}
This statement is a quotation and citation from the BH Complementary paper \cite{Susskind:1993if} and we refer the reader to the original paper for more details. Just to provide a few words here, it is supported by the original semi-classical derivation of the Hawking radiation.  Calculation of the Bogoliubov coefficients requires that we follow outgoing modes backwards in time to a region in the vicinity of the horizon, namely to a cut-off distance $\ell_p$ \cite{Jacobson:1991gr}. This means that as seen by an $r_0$-Fido with $r_0=r_{_\mathrm{H}}+\epsilon$ and $\epsilon\sim \ell_p$, there exists a flux of energy. This energy is carried radially and outward by the Hawking radiation. This radiation is intense, because the frequency $\nu_{r_0}$  is related to the frequency at asymptotics $\nu_\infty$  by the blue-shift factor $\nu_{r_0}=\dfrac{\nu_\infty}{\sqrt{f(r_0)}}$ which grows very fast as $r_0\to  r_{_\mathrm{H}}$. 

For the next step, we focus on a family of Frefos which help us to find the inconsistency. Intuitively, they are radially free-falling observers which are dropped from the radius $r_0$, and we call them $r_0$-Frefo. In this regard, one can imagine an  $r_0$-Frefo as a free-falling observer which has been released from the radius $r_0$ with zero velocity.  The time at which an $r_0$-Frefo can be released is irrelevant in our setup, because eventually we are going to study the Hawking radiation from the BH, which is pretty stable in time scales that we are interested in. Besides, we will study $r_0$-Frefos only at the very beginning of their falling journey, i.e. when they are still close to the radius $r_0$. We consider such Frefos on the Schwarzschild background, and label them by their release radius $r_0$ as $r_0$-Frefo. It is clear that by the spherical symmetry of the background and the radial motion, we have dropped the $(\theta,\varphi)$ coordinates in labeling the $r_0$-Frefos. In the next statement, we compare the frequencies observed by the $r_0$-Fidos and $r_0$-Frefos.
\begin{itemize}
\item [] {\textbf{Statement 3}:} There is no gravitational redshift between an $r_0$-Fido and an $r_0$-Frefo at $r_0$.
\end{itemize}
To clarify this statement, we notice that not only an $r_0$-Frefo coincides with an $r_0$-Fido in radial distance $r_0$, but also has a zero velocity with respect to her/him.  Therefore, the local Lorentz transformation is a trivial one, and there is not any local Doppler shift for the frequencies. Besides, the two observers are at the same position, and there is not any gravitational redshift too. In the literature this is called \emph{Clock Hypothesis} or \emph{Clock Postulate}. It says that for two observers which are at the same position, the rates of clocks depend only on the relative velocity, and not the acceleration. As an intuition, a spaceship which is standing statically at a constant radius from a star (by using some rockets), and an explorer which is dropped from it with zero initial velocity and begins to move towards that star, both perceive the light of the star with the same color at the very beginning of the journey of the explorer. Now we are ready to provide a corollary which can be considered as the most important part of the analysis.
\begin{itemize}
\item [] {\textbf{Corollary}:} An $r_0$-Frefo with the $r_0=r_{_\mathrm{H}}+\epsilon$ and $\epsilon\sim \ell_p$ encounters high frequency Hawking radiation at the $r_0$.
\end{itemize}
This corollary admits the existence of a firewall and is in contradiction with the Postulate 4. At first glance, this corollary seems to be deduced straightforwardly from the Statement 2 and 3: by the Statement 2, an $r_0$-Fido with $r_0$ in the vicinity of the horizon observes high frequency Hawking radiation, and by the Statement 3, the corresponding $r_0$-Frefo observes the same frequency Hawking radiation. But it is not so straightforward to deduce such a result, because at the semi-classical regime there is a possibility which undermines this deduction: by changing the observers, a quantum state with non-zero number of particles can be mapped to the vacuum state, i.e. the state without any particle. In this case, the Statement 3 does not apply anymore, because one cannot attribute any frequency to the vacuum state. The best example is the Unruh effect in the flat space time \cite{Unruh:1976db} (reviewed in the Appendix): although an accelerating observer can be found at the same position and at rest with respect to an inertial observer, but their observation from the Unruh particles are totally different; the accelerating observer detects a thermal bath of a finite and non-zero temperature, while the inertial observer detects nothing, just the vacuum.  Provided that in the $(t,r)$-sector of the  near horizon of a Schwarzschild BH one finds a 2-dimensional Rindler space-time, we need to be careful about such a possibility.

Here, we show that the vacuum is not a possibility for the Frefos which are singled out in the Corollary. The argument is in the setup of the Hawking radiation clarified in the Statements 1 and 2, and is based on a crucial and very important difference between the Hawking radiation and the Unruh effect; in the Hawking radiation, as seen by an $r_0$-Fido there is a non-vanishing flux of energy outward, while for a constant accelerating observer in flat spacetime, the Unruh effect is a \emph{thermal bath}, and there is not a net flux of energy in any directions. The former has been explained in the Statement 2 and we take it as granted by the Postulate 2 (see the next section for more discussion). To clarify the latter, we recall the derivation of the Unruh effect (see the Appendix). In the derivation of the Unruh effect for a constant accelerating observer (in any dimensions), we usually use the Cartesian coordinates and align one of the spatial axes with the motion of the observer. Then, using the Bogoliubov transformations, we relate the expansion modes of the inertial observers to the expansion modes of the accelerating observers. The explicit calculation shows that there is not any difference between the negative and positive momentum modes of the accelerating observer  (see e.g. explicit calculation in \cite{Mukhanov:2007zz,Fabbri} and discussion about this issue there). As a result, there is the same thermal distribution of particles for both of the negative and positive momentum modes in the Rindler coordinates.  We postpone more discussions on this difference, its relation to the topology of the horizons, and assessment of the near horizon limit of the Schwarzschild BH to the next section, and continue the argument based on this difference.  

From the point of view of a Fido, there is a net flux of energy outward, and the interaction of this flux with a Frefo can cause a deviation in geodesic for her/him.  In other words, as seen by a Fido the radiation pressure of the Hawking radiation can change the world-line of a Frefo out of its geodesic.  To facilitate this, the Fidos can use some mirrors to shed more radiation on a Frefo and make the deviation as large as they want. From the point of view of Fidos, this is a similar phenomenon as the solar radiation pressure which may affect the motion of the satellites. However, the deviation of the geodesic motion is independent of the observer, and the Frefo her/himself would observe it. But in the semi-classical regime, a Frefo can attribute the deviation of geodesic to the fields other than gravity, because the metric is considered as a classical field in the background. So, her/his state of quantum field has to be a state other than the vacuum, containing some Hawking particles. We emphasize that such a procedure is not possible in the Unruh setup, because there is not any net flux of energy in the thermal bath which has filled the space of an accelerating observer.       

\begin{figure}[ht!]
\vspace*{0.8cm}
\centering
\captionsetup{width=.8\textwidth}	
\begin{tikzpicture}[scale = 1.5]
	\begin{scope}[shift={(-5.5cm,0cm)},rotate=0]
	\draw[draw=lightgray,very thin] (-1.6,1.6) rectangle (3.1,-1.6);
    \shade[ball color=black,opacity=1] (0,0) circle (1cm);
    \node[inner sep=0pt] (Satellite) at (2.2,0)
    {\includegraphics[width=.06\textwidth]{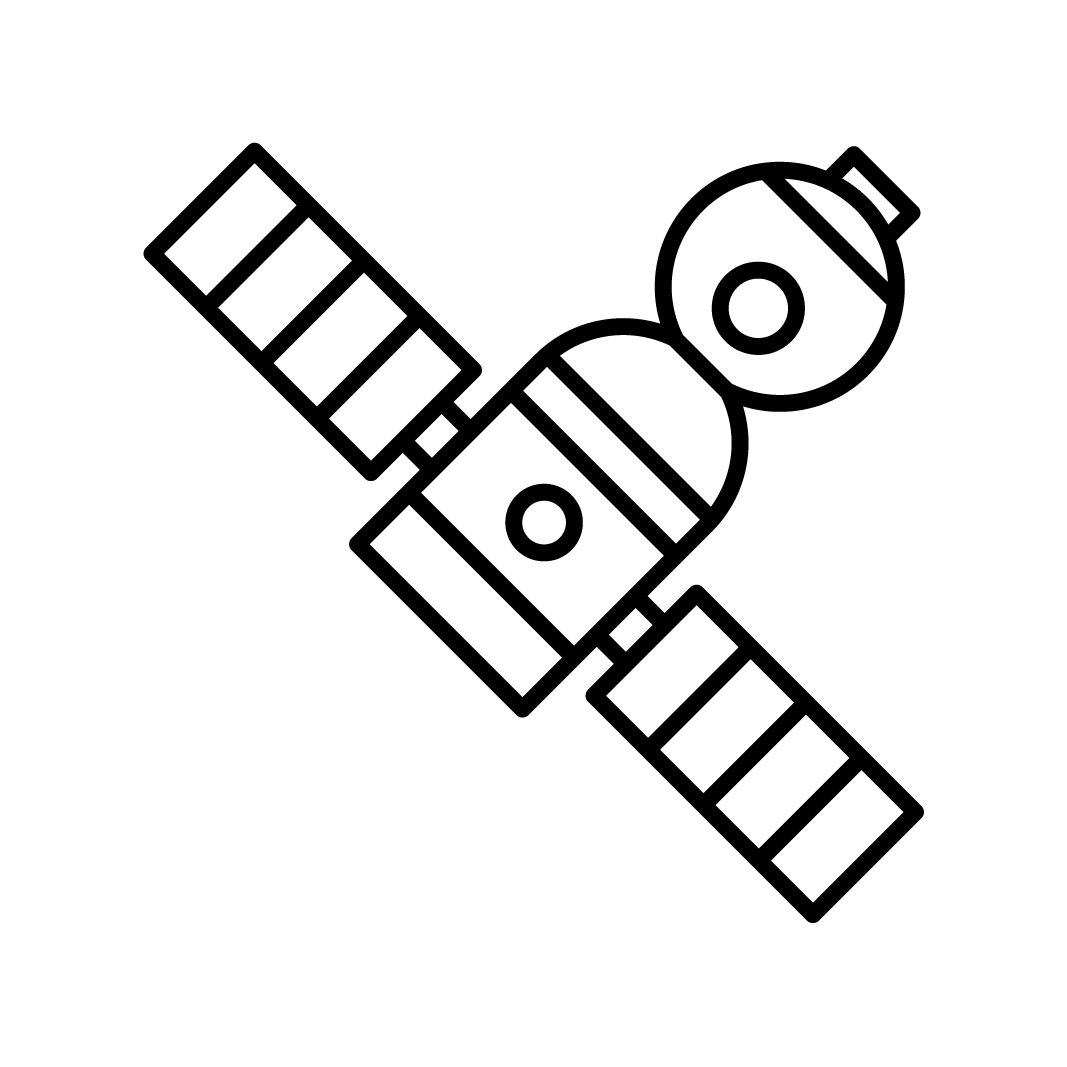}};
    \draw (2.9,-1.3) node[left]{(a)};
	\end{scope}
    \draw[draw=lightgray,very thin] (-1.6,1.6) rectangle (3.1,-1.6);
    \node[inner sep=0pt] (Satellite) at (2.2,0)
    {\includegraphics[width=.06\textwidth]{Satellite.png}};
    \shade[ball color=blue!10!white,opacity=0.20] (0,0) circle (1cm);
    \tikzset{snake it/.style={decorate, decoration=snake,segment length=1.7mm}}
    \draw[thick,->, snake it, rotate around={0:(0,0)}] (0,1.1) -- (0,1.5);
    \draw[thick,->, snake it, rotate around={45:(0,0)}] (0,1.1) -- (0,1.5);
    \draw[thick,->, snake it, rotate around={90:(0,0)}] (0,1.1) -- (0,1.5);  
    \draw[thick,->, snake it, rotate around={135:(0,0)}] (0,1.1) -- (0,1.5); 
    \draw[thick,->, snake it, rotate around={180:(0,0)}] (0,1.1) -- (0,1.5); 
    \draw[thick,->, snake it, rotate around={225:(0,0)}] (0,1.1) -- (0,1.5);
    \draw[thick,->, snake it, rotate around={270:(0,0)}] (0,1.1) -- (0,1.5);
    \draw[thick,->, snake it, rotate around={315:(0,0)}] (0,1.1) -- (0,1.5);
    \draw (2.9,-1.3) node[left]{(b)};
\end{tikzpicture}
\caption*{\footnotesize (a) In the classical regime a black hole does not radiate, and a free-falling object, e.g. a satellite, moves on a geodesic line. (b) In the semi-classical regime, a non-extremal black hole emits Hawking radiation. From the point of view of observers standing outside, the radiation originates from the stretched horizon, and is a flux of energy which affects the motion of the free-falling object. From the viewpoint of the fixed observers this effect grows as the falling object approaches the stretched horizon, because of larger fluxes and highly blue-shifted frequencies. However, the deviation from the geodesic motion is observer-independent, and the observers attached to the free-falling object can detect it, e.g. by experiencing an acceleration in their capsule locally. Encountering such a strong effect while falling towards the stretched horizon is in contradiction with the No Drama postulate. The main point is that in this argument, we use only the Postulates (2) and (4).}
\end{figure}

In the end, we emphasize the following three issues on the derivation of the paradox.
\begin{itemize}
\item [-] The smallness of $\epsilon$ is in correspondence with the qualitative word ``high frequency" in Corollary, and its magnitude does not play an important role in the logical steps of the paradox.
\item [-] It does not matter whether the black hole is eternal, or it has been created from a collapse.
\item [-] As a result of the Corollary and the blueshift of the $r_0$-Frefo's outgoing photons as seen by other infalling Frefos, the firewall paradox is also inevitable for the standard Frefos which are released from a distance.
\end{itemize}

\section{Discussion}

\noindent\underline{Drama for $r_0$-Frefos with arbitrary $r_0$:} One may think about the momentarily stationary motion of the $r_0$-Frefos at the radius $r_0$ as a seemingly problematic issue for the argument: ``if Frefos are momentarily stationary at a Planck distance from a red hot brane, the proximity to the red hot brane burns them, and this is not in contradiction with No Drama." In this scenario, a Frefo which has vanishing relative velocity with respect to the Fidos is burned, while another Frefo who is moving with some velocity towards the BH is not burned!  This argument is a drastic deviation from the equivalence principle; locally in the low curvature regime, special relativity should be achieved which does not distinguish between inertial frames based on their relative velocities. The only concern could be the Doppler effect in which the relative velocity plays a role. However, the Doppler effect acts against the scenario, i.e. if an $r_0$-Frefo is burned in the vicinity of the horizon, then another Frefo who is falling towards the BH with some relative velocity should be burned worse by a blueshifted hot brane. So, this argument does not remove the contradiction, and just transfers it into the question of ``why one of the Frefos is burned, while the other one is not?". 

\noindent\underline{Drama for $r_0$-Frefos irrespective of their past dynamical history:} In the argument of this paper, we prepared an $r_0$-Frefo by moving a distant observer slowly  to the radius $r_0$, and then releasing him/her to fall radially towards the black hole.  However, one could suggest another setup to introduce an $r_0$-Frefo: a freely falling observer who moves radially outward,  reaches to a maximum radius $r_0$, and then falls towards the smaller radii. This motion resembles shooting a ball from the earth vertically and upward with a velocity less than its escape velocity. The ball moves upward, reaches its maximum distance from the earth, and falls down. At the maximum radius (let's call it $r_0$), it resembles an $r_0$-Frefo locally.  In the limit $r_0\gg r_{_\mathrm{H}}$ one recovers the Painlev\'e-Gullstrand observers, i.e. Frefos which are dropped from asymptotics with vanishing initial velocity. In general, it does not matter how an $r_0$-Frefo is prepared. The only relevant properties of this observer are to coincide in position, and be momentarily at zero relative velocity w.r.t. a Fido at the radius $r_0$. This configuration can be prepared in infinitely many different ways, which are irrelevant to the argument. 

\noindent\underline{Past horizon singularity:}  In our setup, the $r_0$ is chosen to be close to a black hole horizon, and one could think that the past horizon can play an important role, and deduce that by the singularity of the Rindler quantum state at the past horizon the argument fails to show the paradox. However, one way to show the irrelevance of the past horizon effect is focusing on a black hole which is created from a collapse.  

\noindent\underline{Near horizon analysis:} By introducing the near horizon radial coordinate $\rho\geq 0$ via $r=r_{_\mathrm{H}}+\frac{\rho^2}{4r_{_\mathrm{H}}}$ and keeping the leading powers of $\rho$, then the near horizon of the Schwarzschild metric \eqref{Schwarzschild} is
\begin{equation}\label{near horizon}
\mrd s^2=\big( -\kappa^2\rho^2\mrd t^2+\mrd \rho^2+r_{_\mathrm{H}}^2(\mrd \theta^2+\sin^2\theta \mrd \varphi^2)\big)\big( 1+\mathcal{O}(\rho^2)\big)
\end{equation}
in which $\kappa=\frac{1}{4M}$ is the surface gravity at the horizon. This is the metric of $\mathbb{R}^{(1,1)}\times S^2$ and is not a flat spacetime in $4$ dimensions. The $\mathbb{R}^{(1,1)}$ sector, i.e. the $-\kappa^2\rho^2\mrd t^2+\mrd \rho^2$ part, is a $(1+1)$ dimensional Minkowski, while the $(t,\rho)$ coordinates cover only the Rindler wedge, in association with the outer side of the BH near horizon region. The bifurcation of the horizon is a two dimensional surface at $\rho=0$ and is compact with the topology of $S^2$ which is parametrized by the coordinates $(\theta, \varphi)$. If we study the wave equation for a scalar field $\Phi(x^\mu)$, i.e. $\Box \Phi=0$, on the $\mathbb{R}^{(1,1)}\times S^2$ background with the metric in the Rindler coordinates \eqref{near horizon} we will find
\begin{equation}\label{split}
 \Big(\frac{-\partial^2_t}{\kappa^2\rho^2}+\partial^2_\rho+\frac{\partial_\rho}{\rho}+\frac{1}{r_{_\mathrm{H}}^2\sin^2\theta}[(\sin\theta\partial_\theta)^2+\partial^2_\varphi]\Big)\Phi=0.
\end{equation}
As a result of the spherical symmetry, the $(t,\rho)$ and $(\theta,\varphi)$ sectors of $\Phi$ separate by the expansion in terms of the spherical harmonics $Y_{l,m}(\theta,\varphi)$
\begin{equation}\label{wave 1}
 \Big(\frac{-\partial^2_t}{\kappa^2\rho^2}+\partial^2_\rho+\frac{\partial_\rho}{\rho}-\frac{l(l+1)}{r_{_\mathrm{H}}^2}\Big)\Phi_l(t,\rho)=0.
\end{equation}
In order to investigate the Hawking radiation, it is usual to focus on the $l=0$ cases for simplicity: 
\begin{equation}\label{wave 2}
 \Big(\frac{-\partial^2_t}{\kappa^2\rho^2}+\partial^2_\rho+\frac{\partial_\rho}{\rho}\Big)\Phi_0(t,\rho)=0.
\end{equation}
This is the wave equation on the 2 dimensional flat metric $-\kappa^2\rho^2\mrd t^2+\mrd \rho^2$.  The calculation in this sector is exactly the same as the Unruh effect in $(1,1)$ dimensional flat spacetime for an accelerating observer with the constant acceleration $a$ such that $\kappa \leftrightarrow a$. In this (reduced) setup, the bifurcation of the horizon is a point in $\mathbb{R}^{(1,1)}$. Based on this analogy, there exists  a thermal bath of the temperature $T=\frac{\kappa}{2\pi}$ in the near horizon region (as seen by the distant observers), in correspondence with the vacuum state of an inertial observer in the $(1+1)$  flat spacetime. This latter is considered to admit the No Drama for Frefos in the near horizon metric.
\begin{itemize}
\item  \textit{Acceleration issue:} As far as the observers are concerned, there is a subtle difference in the  near horizon vs. Rindler analogy. The Unruh effect in the Rindler case maps the vacuum state of an inertial observer to the thermal state of an accelerating observer. So, the price that the latter observer  pays to enjoy having thermal particles is to undergo (not a joyful) acceleration.  However, in the near horizon analysis of the Schwarzschild BH,  both of the observers whose states are mapped to, are inertial: the Frefo in the near horizon region vs. the asymptotic observer. Both of such observers do not feel locally any acceleration or gravitational force.   In other words, the asymptotic observer enjoys having thermal particles with the temperature $T=\frac{\kappa}{2\pi}$, which is analogous to temperature of the Rindler accelerating observer, without feeling any acceleration or gravitational force her/himself. The point is that the asymptotic observer associates a surface gravity to the region around the horizon, i.e. to the Fidos in the near horizon region, not to her/himself. We notice that, by the Statement 2, it is not possible to associate the temperature $T=\frac{\kappa}{2\pi}$ to the observations of a Fido in the vicinity of the horizon, which contradicts the high blue-shift of the Hawking temperature.  
\end{itemize}

\noindent\underline{Importance of the energy flux:} Putting the BH in a box (or in the AdS background) is a way to make the importance of the energy flux in the argument manifest. Such a BH can be in equilibrium with its Hawking radiation, i.e. it can emit and absorb the same amount of radiation. By the time-reversal symmetry one can see that for such a BH in equilibrium with its radiation, the radiation process is unitary. However, with a small amount of energy flux at infinity, the information loss paradox appears in the semi-classical regime. Therefore, the presence of flux results to a paradoxical seemingly non-unitary evaporation process. Our analysis in this paper makes it clear that there is a similar issue in  studying the firewall paradox: for a BH in a box and in the absence of a net flux of energy, the derivation of the firewall paradox (in its modified version in this paper) fails. However, with a small amount of energy flux, the firewall paradox appears. This is an interesting outcome of the analysis which brings the information and firewall paradoxes on the same stage using the importance of the role of the energy flux. Such a strange discontinuous behavior of unitary$\leftrightarrow$non-unitary and no-firewall$\leftrightarrow$firewall for BHs in AdS backgrounds may provide some hints in favor of the unitarity and no-firewall resolutions of the paradoxes.  Notice also that the presence of the energy flux does not depend on the total evaporation process of the BH, and is independent of Postulates 1 and 3. This is manifest by comparing two famous Hawking's papers: one on the presence of the Hawking radiation without considering full evaporation of the BH \cite{Hawking:1976rt}, and the other paper on the information paradox which considers the BH evaporation \cite{Hawking:1976ra} which brings the Postulates 1 and 3 into the scene. However, in both of the papers (hence, in the first paper for our interest here) there exists a net flux of energy via the Hawking radiation.     

\noindent\underline{Emission source of the Hawking radiation:} Although we took the Statement 2 as granted, there are alternative points of view which do not consider the (stretched) horizon as the origin of the Hawking radiation. As an example one can refer to the ``quantum atmosphere" idea in Refs. \cite{Giddings:2015uzr,Dey:2017yez}. However, there remains a question about the possibility of a consistent formulation of the idea such that it respects the semi-classical approximation Postulate 2 manifested in the original Hawking calculation. 

\noindent\underline{A proof of the Firewall?} The steps in this paper which led to the main corollary can be considered as a proof for the existence of the firewall for the Frefos diving into the black hole from outside of the stretched horizon, if one takes Postulate 2 (and Statement 2) as granted. Nonetheless, to be conservative, we did not promote the analysis in this way. From this point of view, one can conclude that the firewall is inevitable, unless one relaxes the semi-classical approximation, the Postulate 2. 

\noindent\underline{What about Postulates 1 and 3?} The result of this paper neither guarantees the correctness of the postulates 1 or 3, nor prohibits other possible contradiction between any set of postulates among the four postulates. However conservatively, if only one of the postulates is going to be relaxed, it should be one of the postulates 2 or 4. \\

\section{Conclusion}
In this paper we showed that the semi-classical approximation and No Drama postulates are in contradiction with each other, by providing an argument (based only on the Postulate 2) in support of the existence of high frequency radiation for Frefos close to the horizon.  The argument consisted of two parts: deducing the existence and deducing high frequencies of radiation for Frefos close to the horizon. For the first part, i.e., the existence argument, we followed these steps:   
\begin{itemize}
\item assuming the existence of Hawking radiation for Fidos (in Statement 1),
\item assuming the flow of energy outward by this radiation as seen by Fidos (in Statement 2),
\item assuming that, from the point of view of Fidos,  this energy flow  exerts radiation pressure on the free-falling objects, and makes some deviation in their geodesic motion (in the derivation of the Corollary),
\end{itemize}
and finally, by observer independence of the deviation from the geodesics (also in the derivation of the Corollary), we deduced that Frefos agree on their non-geodesic motion. However, from their point of view this deviation is only possible if there exist some radiation to interact with.  

The second part, i.e. the high frequency of the Frefos radiation was achieved by: 
\begin{itemize}
\item assuming the high frequencies for the radiation as seen by Fidos in the vicinity of the horizon (in Statement 2),
\item and assuming that the observed frequency of a light beam does not depend on the local acceleration of the momentarily coinciding observers (in Statement 3).
\end{itemize}
This final step implies that a Frefo which momentarily coincide in position with a Fido near the horizon encounters high frequency radiation as well. This result is in tension with the No Drama version of the equivalence principle, which is taken as granted by  the Postulate 4. 

Modification of the steps described above can force us to go beyond the standard physics. For example, proposing a thermal radiation without exertion of pressure, or acceleration dependency of the frequencies of such a radiation would break the locality in physics: local behavior of a beam of radiation should not depend on the source of the emission, whether it is a black hole or it is a shining star.

\appendix

\section{Review of Unruh effect}

Here, we review calculation of the Unruh effect for two purposes: 1) to make it clear that in this  calculation (and similarly in the derivation of Hawking radiation), the postulates 1 and 3 are not assumed, 2) to show explicitly that the Unruh effect is a thermal bath, in contrast with the Hawking radiation which is an outgoing radiation. For simplicity, we consider the  setup in 2-dimensional space-time. An interested reader can find more details in e.g. \cite{Mukhanov:2007zz}. 

One begins with the flat metric for the inertial observers $\mrd s^2=-\mrd t ^2 +\mrd x^2$ in the coordinates $(t,x)$. For an accelerating observer with an acceleration $a$ the metric is $\mrd s^2=e^{2a\xi}(-\mrd \tau^2+\mrd \xi^2)$ in the coordinates $(\tau,\xi)$ . The two set of coordinates are related to each other by
\begin{equation}
t=\frac{1}{a}e^{a\xi}\sinh (a\tau), \qquad x=\frac{1}{a}e^{a\xi}\cosh (a\tau).
\end{equation}

In inertial frames, i.e. the $(t,x)$ coordinates, equation of motion for a real scalar field $\Box \Phi =0$ is simply $(-\partial^2_t+\partial_x^2)\Phi=0$. Classical solutions to this equation are expanded as the following:
\begin{equation}
\Phi(t,x)=\int_{-\infty}^{\infty}\frac{\mrd k}{\sqrt{2\pi}} \frac{1}{\sqrt{2\omega_{_k}}}\Big(a_{_k} e^{-i\omega_k t+ikx} + a^{\dagger}_{_k} e^{i\omega_k t-ikx} \Big), \qquad \omega_{_k}\equiv |k|,
\end{equation}
which equivalently can be written as
\begin{align}
\Phi&=\int_{0}^{\infty}\frac{\mrd \omega}{\sqrt{2\pi}} \frac{1}{\sqrt{2\omega}}\Big(a_{_\omega} e^{-i\omega(t-x)} + a_{_{-\omega}} e^{-i\omega (t+x)}+a^\dagger_{_\omega} e^{i\omega(t-x)} + a^{\dagger}_{_{-\omega}} e^{i\omega (t+x)}\Big)\\
& =\int_{0}^{\infty}\frac{\mrd \omega}{\sqrt{2\pi}} \frac{1}{\sqrt{2\omega}}\Big(a_{_\omega} e^{-i\omega \bar u} + a_{_{-\omega}} e^{-i\omega \bar v}+a^\dagger_{_\omega} e^{i\omega\bar u} + a^{\dagger}_{_{-\omega}} e^{i\omega \bar v}\Big). \label{expansion 1}
\end{align}
In the last equation, the light-cone coordinates $\bar u = t-x$ and $\bar v= t+x$ are introduced.  So, $\Phi=A(\bar u)+B(\bar v)$ and 
\begin{align}\label{A B}
A(\bar u)=\int_{0}^{\infty}\!\!\frac{\mrd \omega}{\sqrt{2\pi}} \frac{1}{\sqrt{2\omega}}\Big(a_{_\omega} e^{-i\omega \bar u} +a^\dagger_{_\omega} e^{i\omega\bar u} \Big), \quad B(\bar v)= \int_{0}^{\infty}\!\!\frac{\mrd \omega}{\sqrt{2\pi}} \frac{1}{\sqrt{2\omega}}\Big( a_{_{-\omega}} e^{-i\omega \bar v}+ a^{\dagger}_{_{-\omega}} e^{i\omega \bar v}\Big).
\end{align}
In the canonical quantization, the complex numbers $a_{_\omega}$ and $a_{_{-\omega}}$ are replaced with operators, and we will not use a new notation for them.

For the accelerating observers, i.e. coordinates $(\tau,\xi)$, the $\Box \Phi =0$ is a similar equation $(-\partial^2_\tau+\partial_\xi^2)\Phi=0$. So, we do not need to repeat the calculations, and simply replace $t\to \tau$ and $x \to \xi$ in the equations above, with the notation $\omega \to \Omega$ and $a_{_\omega} \to b_{_\Omega}$. Therefore, for the accelerating observers we find $\Phi=P(u)+Q(v)$ in which 
\begin{align}\label{P Q}
P(u)=\int_{0}^{\infty}\!\!\frac{\mrd \Omega}{\sqrt{2\pi}} \frac{1}{\sqrt{2\Omega}}\Big(b_{_\Omega} e^{-i\Omega u} +b^\dagger_{_\Omega} e^{i\Omega u} \Big), \quad Q(v)= \int_{0}^{\infty}\!\!\frac{\mrd \Omega}{\sqrt{2\pi}} \frac{1}{\sqrt{2\Omega}}\Big( b_{_{-\Omega}} e^{-i\Omega  v}+ b^{\dagger}_{_{-\Omega}} e^{i\Omega  v}\Big).
\end{align}
The relation of the light-cone coordinates of the accelerating observers $u=\tau-\xi$ and $v=\tau+\xi$ and the ones for the inertial frame $(\bar u, \bar v)$ is
\begin{equation}\label{u ubar}
\bar u= \frac{-e^{-au}}{a}, \qquad \bar v= \frac{e^{av}}{a}.
\end{equation}
In Bogoliubov method \cite{Bogoliubov}, the goal is to find the relations between the basis $a_{_\omega}$ and $b_{_\Omega}$. To be more specific, one tries to find the coefficients $\alpha_{_{\omega\Omega}}$ and $\beta_{_{\omega\Omega}}$ such that
\begin{equation}\label{Bogoliubov def}
b_{_\Omega}=\int_0^\infty \mrd \omega (\alpha_{_{\omega\Omega}}a_{_\omega}+\beta_{_{\omega\Omega}}a^\dagger_\omega).
\end{equation} 
To this end, one can use $\Phi(u, v)=\Phi(\bar u, \bar v)$ for the corresponding coordinates (via \eqref{u ubar}), which yields
\begin{equation} \label{P=A}
P(u)=A(\bar u), \qquad Q (v)=B(\bar v). 
\end{equation}
First we look at the first equation of these two, which is associated with the right moving modes, by taking Fourier transform of both sides
\begin{equation}\label{LHS RHS}
\int_{-\infty}^\infty \frac{\mrd u}{\sqrt{2\pi}} e^{i\Omega u} P(u)= \int_{-\infty}^\infty \frac{\mrd u}{\sqrt{2\pi}} e^{i\Omega u} A(\bar u). 
\end{equation} 
Replacing $P$ from \eqref{P Q}, the l.h.s is 
\begin{align}
\int_{0}^{\infty}\!\!\frac{\mrd \Omega'}{\sqrt{2\pi}} \frac{1}{\sqrt{2\Omega'}} \int_{-\infty}^\infty \frac{\mrd u}{\sqrt{2\pi}}& e^{i\Omega u} \Big(b_{_\Omega} e^{-i\Omega' u} +b^\dagger_{_\Omega} e^{i\Omega' u} \Big)\\
&= \int_{0}^{\infty}\!\!\frac{\mrd \Omega'}{\sqrt{2\pi}} \frac{1}{\sqrt{2\Omega'}} \Big(b_{_\Omega} \delta(\Omega-\Omega') +b^\dagger_{_\Omega} \delta (\Omega+\Omega') \Big) = \frac{1}{\sqrt{2\Omega}}b_{_\Omega}, \label{final lhs}
\end{align}
in which the definition of the Dirac delta function $\delta (\Omega-\Omega')=\int_{-\infty}^\infty \frac{\mrd u}{\sqrt{2\pi}} e^{i(\Omega -\Omega')u}$ is used. For the r.h.s of \eqref{LHS RHS}, we replace the $A(\bar u)$ from \eqref{A B}  
\begin{align}\label{final rhs}
\int_{0}^{\infty}\!\!\frac{\mrd \omega}{\sqrt{2\pi}} \frac{1}{\sqrt{2\omega}} \int_{-\infty}^\infty \!\!\frac{\mrd u}{\sqrt{2\pi}} e^{i\Omega u} \Big(a_{_\omega} e^{-i\omega \bar u} +a^\dagger_{_\omega} e^{i\omega\bar u} \Big)
= \int_{0}^{\infty}\!\!\frac{\mrd \omega}{\sqrt{2\pi}} \frac{1}{\sqrt{2\omega}} \Big(F(\omega,\Omega) a_{_\omega}  + F(-\omega,\Omega) a^\dagger_{_\omega}  \Big)
\end{align}
in which we defined an auxiliary function
\begin{equation}\label{F def}
F(\omega,\Omega)\equiv  \int_{-\infty}^\infty \!\!\frac{\mrd u}{\sqrt{2\pi}} e^{i\Omega u-i\omega \bar u}= \int_{-\infty}^\infty \!\!\frac{\mrd u}{\sqrt{2\pi}} e^{i\Omega u+ \frac{i\omega e^{-au}}{a}}.  
\end{equation}
According to \eqref{LHS RHS}, the final results in \eqref{final lhs} and \eqref{final rhs} are equal:
\begin{equation}\label{b F omega}
\frac{1}{\sqrt{2\Omega}}b_{_\Omega}= \int_{0}^{\infty}\!\!\frac{\mrd \omega}{\sqrt{2\pi}} \frac{1}{\sqrt{2\omega}} \Big(F(\omega,\Omega) a_{_\omega}  + F(-\omega,\Omega) a^\dagger_{_\omega}  \Big). 
\end{equation}
Comparing this result with \eqref{Bogoliubov def}, the Bogoliubov coefficients are found to be
\begin{equation}\label{alpha beta}
\alpha_{_{\omega\Omega}}=\sqrt{\frac{\Omega}{\omega}}F(\omega,\Omega) \qquad \beta_{_{\omega\Omega}}=\sqrt{\frac{\Omega}{\omega}}F(-\omega,\Omega).
\end{equation} 
Now, we can find the particle number density as seen by an accelerating observer for the vacuum state of an inertial observer. The vacuum state of the inertial observer is denoted by $|0\rangle_{_a}$ which is annihilated by the $a_{_\omega}$ operators, i.e. $a_{_\omega}|0\rangle_{_a}=0$. The expectation value of the accelerating observer's particle number operator
$b^\dagger_{_\Omega}b_{_\Omega}$ in this state is 
\begin{align}
_{_a}\!\langle 0|b^\dagger_{_\Omega}b_{_\Omega}|0\rangle_{_a}&=\int_{0}^{\infty}{\mrd \omega} \beta_{_{\omega\Omega}} ^\ast \beta_{_{\omega\Omega}} = \int_{0}^{\infty}{\mrd \omega} \ {\frac{\Omega}{\omega}}\ F^\ast (-\omega,\Omega) F (-\omega,\Omega)\\
& = \frac{1}{e^{\frac{2\pi\Omega}{a}}-1} \delta (0).\label{Planck dest}
\end{align} 
In the first equality, $b_{_\Omega}$ is replaced from \eqref{Bogoliubov def} and $a_{_\omega}|0\rangle_{_a}=0$ is used. In the second equality, $\beta_{_{\omega\Omega}}$ is replaced by \eqref{alpha beta}. The last equality, i.e. \eqref{Planck dest} is a mathematical relation, and we will prove it at the end of this appendix.  The $\delta (0)$ indicates the divergence of the particle number density which is calculated all over the space. The rest of \eqref{Planck dest} is the bosonic distribution of particles in the temperature $T=\frac{a}{2\pi}$, which is the celebrated Unruh temperature. From the calculations above we notice that:
\begin{itemize}
	\item one can repeat the analysis for the second equation in \eqref{P=A}, i.e. for the left moving modes, and end up with the same result as in \eqref{Planck dest}. In this regard, the left and right moving particle distributions are thermal with the same temperature, and the accelerating observer detects a thermal bath. This result is different from the Hawking radiation emanating from a black hole; the Hawking radiation is a thermal distribution which (in the setup of a collapsing matter) can be calculated only for the outgoing modes.
	
	\item In the calculation above, it is explicit that the unitarity of quantum evolution is not used as an assumption. The canonical quantization procedure deals with construction of a Hilbert space at a given time slice, while the unitarity is a relation between Hilbert spaces at different time slices. So, the unitarity is an extra assumption in the quantization of fields, and explicitly is not used in the calculations above.  Besides, the microstates/entropy of the system does not play any role in the calculations. Considering that the calculation of the Hawking radiation is basically a similar computation in the near horizon region, we deduce Statement 1, i.e. the independence of the existence of Hawking radiation from the Postulates 1 and 3.  
\end{itemize}
 
\noindent\underline{Proof of \eqref{Planck dest}}: As one of the necessary conditions for the invertibility of Bogoliubov transformation between the two bases (not to be confused with the unitarity, or the invertibility of the evolution of the system which is related to the norm of the state of the system), we have $\alpha^\dagger \alpha - \beta^\dagger \beta = I $, which in our set up it is:
\begin{align}
\int_{0}^{\infty}{\mrd \omega} (\alpha^\ast_{_{\omega\Omega'}}\alpha_{_{\omega\Omega}}-\beta^\ast_{_{\omega\Omega'}}\beta_{_{\omega\Omega}} )= \delta (\Omega-\Omega').
\end{align}
Replacing \eqref{alpha beta} in the l.h.s we find
\begin{align}
\int_{0}^{\infty}{\mrd \omega}\frac{\sqrt{\Omega\Omega'}}{\omega} \left( F^\ast (\omega,\Omega')F(\omega,\Omega)-F^\ast (-\omega,\Omega')F (-\omega,\Omega)  \right)= \delta (\Omega-\Omega').
\end{align}
Using the relation $F (\omega,\Omega)=e^{\frac{\pi \Omega}{a}}F (-\omega,\Omega)$ (which is found from \eqref{F def} via change of variable $u=u'-\frac{i\pi}{a}$) in this equation:
\begin{align}
\int_{0}^{\infty}{\mrd \omega}\frac{\sqrt{\Omega\Omega'}}{\omega}\left( e^{\frac{\pi (\Omega+\Omega')}{a}}-1 \right)F^\ast (-\omega,\Omega')F (-\omega,\Omega)= \delta (\Omega-\Omega').
\end{align} 
Finally, setting $\Omega=\Omega'$ yields the \eqref{Planck dest}, and derivation of the Unruh effect is completed.

\noindent \textbf{Acknowledgements:} I acknowledge the comments and contributions from Volker Perlick to this paper, which improved the arguments very much. Besides, I am very grateful for the kind supports from Jutta Kunz in Oldenburg University. I would also like to thank Niayesh Afshordi, Furkan Semih Dündar, Daniel Grumiller, and Shahin Sheikh-Jabbari for useful discussions. Moreover, I thank the anonymous referee who pointed out the importance of avoiding the past horizon singularity in the argument. This work has been supported by TÜBITAK international researchers program No. 2221.

\end{document}